\documentclass[aps, pra, twocolumn, notitlepage, superscriptaddress, nofootinbib   
]{revtex4-1}
\usepackage{amssymb,latexsym,amsmath}
\usepackage{bm}
\usepackage{amssymb,amsmath}
\usepackage{graphicx,xcolor}

\usepackage[caption=false,lofdepth,lotdepth]{subfig}

\usepackage{hyperref}   
\hypersetup{%
	pdfpagemode=UseNone, 
	pdfstartpage=1,
	pdfmenubar=true,
	pdftoolbar=true,
	colorlinks = true,
	linkcolor=blue,
	citecolor=blue,
	urlcolor=blue,
	bookmarksopen=false
}

\newcommand{\beq}{\begin{equation}}   
\newcommand{\eeq}{\end{equation}}

\begin{document}
	
\title{Superfluid vortex dynamics on an ellipsoid and other surfaces of revolution}
\author {M\^onica A.~Caracanhas}
\affiliation{Instituto de F\'isica de S\~ao Carlos, Universidade de S\~ao Paulo, S\~ao Paulo, Brazil}
\author {Pietro Massignan}
\email{pietro.massignan@upc.edu}
\affiliation{Departament de F\'isica, Universitat Polit\`ecnica de Catalunya, Campus Nord B4-B5, E-08034 Barcelona, Spain}
\author {Alexander L.\ Fetter}
\email{fetter@stanford.edu}
\affiliation {Departments of Physics and Applied Physics, Stanford University, Stanford, CA 94305-4045, USA}
\date{\today}

\begin{abstract}

We study the dynamics of quantized superfluid vortices on axisymmetric compact surfaces with no  holes, where the total vortex charge must vanish and the condition of irrotational flow forbids  distributed vorticity. A conformal transformation from the surface to the complex plane allows  us to use familiar formalism to describe the motion of the quantized vortices and to find the total energy.  The simplest case is a vortex dipole with unit vortex charges on an axisymmetric ellipsoid.  We study two special symmetric vortex-dipole configurations along with a general asymmetric one. 

\end{abstract}

\maketitle

\section{Introduction}

Quantized vortices play an essential role in the physics of superfluids, such as $^4$He-II~\cite{Donnelly1991book}, Bose-Einstein condensates (BECs), and fermionic superfluids of ultracold dilute atomic gases~\cite{PethickSmith2008book, PitaevskiiStringari2016book}.  The dynamics of quantized vortices simplifies considerably for thin films where vortex-line bending is absent.
At low temperature, these superfluids are effectively ideal fluids with no dissipation and  obey equations of classical hydrodynamics~\cite{Kirchhoff1876,Lamb1945}, along with the condition of quantized vorticity~\cite{Onsager1949,Feynman1955}.  

Most studies have focused on planar superfluid films, but Lamb remarks in Sec.~160 of \cite{Lamb1945} that the theory can be generalized to curved surfaces and quotes some results for a spherical film. He relies on the method of stereographic projection,
which  maps  the  surface of a sphere to the complex plane where  the familiar formalism of vortex hydrodynamics works well. A recent work by two of us~\cite{Bereta2021} provides more details on this approach on a sphere along with many other references. Other simple conformal transformations permit studying the dynamics on surfaces with zero Gaussian curvature, such as cylinders and cones \cite{Ho2015,Guenther2017,Massignan2019}.
 
Many theoretical papers have  explored the properties of a BEC trapped in a spherically symmetric shell potential \cite{Sun2018, Zhang2018, Bereta2019, Bereta2021, Tononi2019, Tononi2021},  but recent 
 experimental advances raise  the important new question of vortex dynamics on axisymmetric ellipsoidal surfaces.
   Indeed, shell potentials for ultracold atoms have been generated by superposing 
quadrupolar and oscillatory magnetic fields~\cite{Lundblad2019}. The resulting spatially dependent dressed atomic states experience an effective 
 axisymmetric ellipsoidal potential.
 
Experiments performed in the NASA Cold Atom Lab aboard the International Space Station recently produced three-dimensional BECs with no gravity~\cite{Aveline2020}, providing now a more realistic possibility to explore bubble trapped BECs~\cite{Carollo2021}. 
One of the aims of these experiments is the generation of vortices, which can be 
achieved by rotating the dressed trap or 
through the spontaneous creation of vortex-antivortex pairs across the condensation transition, namely through the Kibble-Zurek mechanism~\cite{Freilich2010,Padavi2020,Carollo2021}.

In contrast to the spherical case, however, finding a conformal projection of an ellipsoidal surface to the flat plane is a complicated task. To solve this problem, 
we rely 
on the method of isothermal coordinates, inspired by Kirchhoff's early study of the surface of a torus \cite{Kirchhoff1875,Kirchhoff1876,Guenther2020}.  Briefly, one seeks a coordinate transformation of the relevant surface metric to make it resemble the metric on the complex plane.  For a sphere, this method immediately reproduces the result of stereographic projection, but it applies much more widely.  Section \ref{sec:formalism} uses this metric method for quite general axisymmetric compact surfaces, and we illustrate it with the complex potential for a vortex dipole on one such surface, including the phase pattern and streamlines. {Reference~\cite{Dritschel2015} anticipated several of our results in  more formal mathematical language.  See also Refs.~\cite{Hally1980,Castilho2008,Regis2018} for other related material.} Section \ref{sec:dynamics} then studies  vortex dynamics on such a general axisymmetric surface.  Similar to the situation on a plane, these dynamical equations assume a Hamiltonian form with the coordinates of each vortex serving as a pair of canonical  variables and the total energy serving as the Hamiltonian.  One new feature is the appearance of local curvature contributions to the energy \cite{Turner2010,Guenther2020} that play an essential role in the vortex dynamics.  In Sec.~\ref{sec:ellipsoid} we apply this general  formalism to an ellipsoid of revolution.  Section \ref{sec:dynamicsOfDipoles} studies the dynamics of a vortex dipole  for two highly symmetric initial configurations on an axisymmetric ellipsoid, along with a more general asymmetric configuration.  We end with conclusions and discussion in Sec.~\ref{sec:conclusions}.
                                               
\section{Axisymmetric compact surfaces} 
\label{sec:formalism}
The surface of a sphere of radius $a$  satisfies the familiar set of equations
\begin{equation}\label{sphere}
x = a\sin\theta\cos\phi,\quad y = a\sin\theta\sin\phi,\quad z =  a\cos\theta,
\end{equation}
where ($\theta,\phi$) are the usual spherical polar coordinates, with $0<\theta<\pi$ measured from the north pole and $0<\phi< 2\pi$ measured from the $x$ axis.

More generally, the surface of an ellipsoid of revolution has the corresponding set of equations
\begin{equation}\label{ellipsoid}
x = a\sin\theta\cos\phi,\quad y = a\sin\theta\sin\phi,\quad z =  b\cos\theta,
\end{equation}
where $a$ is the radius in the $xy$ plane and $b$ is the radius along the symmetry axis.  In particular, this surface obeys the expected equation
\beq\label{ellipse}
\frac{x^2+y^2}{a^2} + \frac{z^2}{b^2} = 1.
\eeq
Note that $\theta$ is now merely an abstract parameter defining the coordinates of the ellipsoidal surface.  Its  relation to the usual polar angle of spherical coordinates, and more generally to the so-called ``geocentric coordinates'', requires a separate discussion (see Sec.~\ref{sec:ellipsoid}).  

Although we shall specialize to an axisymmetric ellipsoid, the formalism developed below applies widely, and we have found it helpful to consider a  general  compact surface of revolution described by the set of equations {(compare Refs.~\cite{Dritschel2015,Guenther2020})}
\begin{equation}\label{general}
x =f_1(\theta)\cos\phi,\quad y = f_1(\theta)\sin\phi,\quad z =  f_2(\theta),
\end{equation}
{where, as in Eq.~(\ref{ellipsoid}) above, $f_1$ and $f_2$ have the dimension of a length.}
We require $f_1(\theta)> 0 $ to avoid self-intersection,  $f_1(0)=f_1(\pi) = 0$ for a closed surface, and $f_2(0) > 0 > f_2(\pi)$ in analogy with the spherical case.

It  is straightforward to find the metric for the surface defined in Eq.~(\ref{general}):
\beq\label{genmetric}
ds^2 = \left[f_1'(\theta)^2 + f_2'(\theta)^2\right] d\theta^2 + f_1(\theta)^2 d\phi^2  \equiv h_\theta^2 d\theta^2 + h_\phi^2 d\phi^2,
\eeq
where we define the metric parameters
\beq\label{metric}
h_\theta =\sqrt{f_1'(\theta)^2 + f_2'(\theta)^2},\quad h_\phi = f_1(\theta).
\eeq
The presence of two different  parameters $h_\theta$ and $h_\phi$ {in Eq.~(\ref{genmetric})} precludes a simple  description with a single complex variable involving $\theta$ and $\phi$ directly.  

The standard procedure is to seek a coordinate transformation from Eq.~(\ref{genmetric}) to  new ``isothermal'' variables ($u,v$), defined as having an isotropic metric $ds^2 =\lambda^2(du^2+dv^2)$ with an overall scale factor $\lambda$.   For the present study, we 
prefer the polar form $u+iv = \rho\,e^{i\phi}$, with $\rho^2 =u^2 +v^2$, $\tan\phi = v/u$, and  
\beq\label{planemetric} 
ds^2 = \lambda^2 (d\rho^2 +\rho^2d\phi^2).
\eeq

Focus on the common azimuthal angle $\phi$ and write Eq.~(\ref{genmetric}) as
\beq
ds^2 =  h_\phi^2 \left(\frac{h_\theta^2}{h_\phi^2} d\theta^2 + d\phi^2\right).
\eeq
{We now compare this metric to  Eq.~(\ref{planemetric}) for a plane} rewritten in the form 
\beq
ds^2 = \lambda^2\rho^2\left(\frac{d\rho^2}{\rho^2} + d\phi^2\right).
\eeq
Evidently, we should require
\beq\label{integral}
\frac{d\rho}{\rho} = \frac{h_\theta}{h_\phi}\,d\theta,
\eeq
which can be integrated to give
\beq\label{rhoint}
    \ln\rho(\theta) =\int d\theta\,\frac{h_\theta}{h_\phi} = \int d\theta\,\frac{\sqrt{f_1'(\theta)^2 + f_2'(\theta)^2}}{f_1(\theta)},
\eeq
apart from an additive constant that will be irrelevant because of overall vortex-charge neutrality for a compact surface.
As a result, the transformed metric has the desired isothermal form {in Eq.~(\ref{planemetric})}
with  overall scale factor
\beq\label{lambda}
\lambda(\theta) = \frac{h_\phi(\theta)}{\rho(\theta)}.
\eeq 

\section{Vortex dynamics on an axisymmetric compact surface}
\label{sec:dynamics}

An ideal fluid is both irrotational and incompressible,  so that its velocity field $\bm v$ obeys  $\bm\nabla \times \bm v = 0$ and $\bm \nabla~\cdot~\bm v~=~0$.  Remarkably, low-temperature superfluid $^4$He and ultracold dilute superfluid quantum gases both act like ideal fluids, facilitating the  study of these important mathematical models.  These quantum superfluids have a complex  order parameter $\Psi =|\Psi|e^{i\Phi}$  with magnitude $|\Psi|$ and phase $\Phi$. For such one-component quantum systems, the gradient of the phase determines the flow velocity \cite{ Feynman1955} 
\beq\label {irrot}
\bm v = \frac{\hbar}{M}\bm\nabla\Phi,
\eeq
 where $2\pi\hbar$ is Planck's constant and $M$ is the relevant atomic mass.    Equation~(\ref{irrot}) ensures that the flow velocity is irrotational;  more unusually,
  it also implies that the circulation  $\kappa = \oint_{\cal C} d\bm l\cdot \bm v = q\,2\pi \hbar/M$  along any closed path $\cal C$ is quantized in units of $2\pi\hbar/M$, where $q$ is an integer.
  In most cases,  quantized vortices are singly quantized, with unit vortex charge  $q = \pm 1$.
  
  In addition to the representation of $\bm v$ in terms of the quantum velocity potential $\Phi(x,y)$,  the condition  of  incompressibility allows a different representation of $\bm v$  in terms of a vector potential  $\bm A$, with $\bm v = \bm \nabla\times \bm A$.  For a uniform two-dimensional fluid on a plane, this condition simplifies to 
  \beq\label{incom}
  \bm v=\frac{\hbar}{M} \,\hat{\bm n}\times \bm\nabla \chi,
  \eeq
  where $\chi(x,y)$ is the stream function and $\hat{\bm n} = \hat{\bm x}\times\hat{\bm y}$ is the unit normal vector.
  
 Equations (\ref{irrot}) and (\ref{incom}) each
  express the velocity components $v_x$ and $v_y$  as  derivatives of the scalar functions $\chi$ and $\Phi$.  In detail, these equations are the Cauchy-Riemann equations for the complex potential 
 \beq\label{F}
 F(z) = \chi(\bm r) + i\,\Phi(\bm r)
 \eeq
 that depends only on  the single complex variable $z = x+iy$.
  In addition,  the derivative of the complex potential determines the hydrodynamic flow velocity  through the important relation
  \beq\label{Fprime}
  v_y + i v_x= \frac{\hbar}{M}\,\frac{dF}{dz} =\frac{\hbar}{M}\, F'(z).
  \eeq
  
  On the complex plane, familiar  formalism gives the complex potential 
  \beq\label{F0}
  F_0(z) = q_0\ln (z-z_0)
  \eeq
   for a single vortex with charge $q_0$ at $z_0$.   Correspondingly, the complex hydrodynamic velocity field is 
   \beq
   v_y + iv_x = \frac{\hbar q_0}{M(z-z_0)}
   \eeq 
   representing circular flow centered at $z_0$.
   
    On a 
plane, a closed curve $\cal C$  defines an inside and an outside.  The circulation integral  along the contour $\cal C$ surrounding the interior region  characterizes the enclosed vorticity, which  must be an integer multiple of $2\pi\hbar/M$ for a superfluid.  This result follows because Eq.~(\ref{irrot}) ensures that the line integral is essentially the net change in $\Phi$ around $\cal C$.  Each vortex has lines of constant phase emerging from or ending at its location, just like the lines of electric field around point charges in two dimensions.  On a  plane, these lines of constant phase can extend to infinity.

The situation is very different for a compact  surface like that 
introduced in Sec.~\ref{sec:formalism}.  Here, a closed curve merely divides the surface into two  regions.  For the first region, the line integral $\oint_{\cal C} d\bm l\cdot\bm v$ still equals an integral multiple of $2\pi\hbar/M$, but the corresponding lines of constant phase must reconverge somewhere on the compact surface. Specifically, consider the other region, where the contour integral  $\oint_{\cal C} d\bm l\cdot\bm v$ now has an opposite orientation, yielding a circulation equal and opposite to the one in the first region.  As a result, the total vortex charge on any compact surface must vanish, with a vortex dipole as the simplest such configuration.

\begin{figure}[ht]
	\includegraphics[width=0.5\columnwidth]{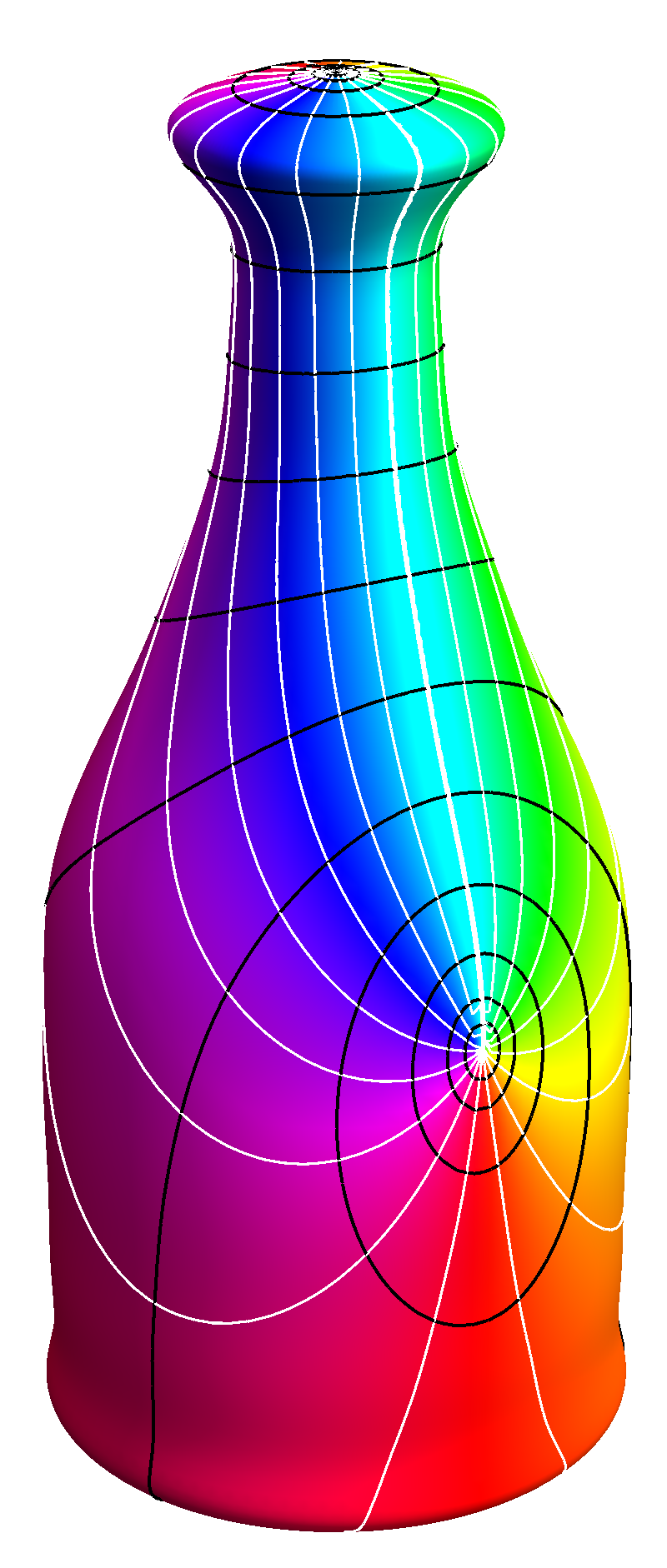}
	\caption{{Color wheel shows} phase pattern generated by a vortex dipole on a superfluid film covering a surface of revolution akin to a champagne bottle. White lines are lines of constant phase and black lines are stream lines. In the configuration displayed, one vortex is at the north pole and the other is on the side.}
	\label{fig:vortexOnABottle}
\end{figure}

For a plane, Eq.~(\ref{F0}) shows that the complex potential for a vortex dipole is the difference of two logarithms $F_{\rm dipole}(z) = \ln(z-z_+)-\ln(z-z_-)=\ln[(z-z_+)/(z-z_-)]$.  The previous coordinate transformation now gives the desired expression for a vortex dipole  on a  quite general  axisymmetric  compact surface
\beq \label{Fdipole}
F_{\rm dipole} =\ln\left(\frac{\rho \,e^{i\phi} -\rho_+\,e^{i\phi_+}}{\rho \,e^{i\phi} -\rho_-\,e^{i\phi_-}}\right)
\eeq
in an obvious short-hand notation.  This complex function gives all necessary information about the hydrodynamic flow field on the surface.  For example, lines of constant phase $\Phi_{\rm dipole} = {\rm Im} \,F_{\rm dipole}$ emerge from the positive vortex and end on the negative vortex.  In addition, lines of constant stream function 
\begin{eqnarray}\label{chidipole}
\chi_{\rm dipole} &=&{\rm Re}\,F_{\rm dipole}\nonumber\\[.2cm]
&=& \frac{1}{2}\ln\left[ \frac{\rho^2 - 2\rho\rho_+\cos(\phi -\phi_+) + \rho_+^2}{\rho^2 - 2\rho\rho_-\cos(\phi -\phi_-) + \rho_-^2}\right]
\end{eqnarray}
trace out the lines of particle flow.  Figure \ref{fig:vortexOnABottle} shows the phase pattern and streamlines for one compact  surface that resembles a champagne bottle, where the function $\rho(\theta)$ was evaluated numerically.

On a planar surface, each member of a vortex dipole moves with the local velocity after subtracting the local circulating velocity of the particular vortex in question. 
But the transformation from the curved surface to the complex plane introduces an additional contribution.  On the physical surface of the object, all point vortices have the same core size $\xi_0$ that serves to regularize the singularity at its center.  On the complex plane,  however, the scale factor $\lambda$ renormalizes each vortex core size to $\xi_j =\xi_0/\lambda_j$, where {$\lambda_j$ denotes the scale factor evaluated at  the position of the $j$th vortex.} See Sec.~VI of Ref.~\cite{Turner2010} for a more detailed discussion of this important issue.

The numerator of Eq.~(\ref{chidipole}) provides the flow arising from the positive vortex.  It contains the combination $\rho^2-2\rho\rho_+\cos(\phi-\phi_+) +\rho_+^2$, which can be rewritten exactly as $(\rho-\rho_+)^2 + 4\rho\rho_+\sin^2[(\phi-\phi_+)/2]$.  For small $d\theta=\theta-\theta_+$ and $d\phi=\phi-\phi_+$, this quantity reduces to {$(\rho_+^2/h_{\phi_+}^2)(h_{\theta_+}^2d\theta^2 + h_{\phi_+}^2d\phi^2) = \rho_+^2\,ds_+^2/h_{\phi_+}^2$.}  In this way, the spatial dependence of the numerator of Eq.~(\ref{chidipole}) becomes $\ln ds_+$, apart from constants that do not affect the hydrodynamic flow.  This term is just the stream function for the circulating flow around the positive vortex.
As noted above, however, this logarithmic singularity must be cut off at the vortex core radius $\xi_+ =\xi_0/\lambda_+$. Hence the dynamics of the positive vortex arises not only from the remaining part of $\chi_{\rm dipole}$ but also from the local curvature involving $\ln\lambda_+$.

The details are somewhat intricate (see Eqs.~(35) and (A7) of~\cite{Guenther2020}) but the final result {for the velocity of the positive vortex} is simple:
\begin{equation}\label{rpdot}
\dot{\bm r}_+ = -\frac{\hbar}{M}\,\hat{\bm n}_+\times \bm \nabla_+\left(\chi_{+-} +\frac{1}{2}\ln\lambda_+\right),
\end{equation}
along with a similar expression for $\dot{\bm r}_-$.  Here $\hat{\bm n}_+$ is the unit normal vector at the position $\bm r_+$ of the positive vortex,  Eq.~(\ref{lambda}) gives the scale factor $\lambda = h_\phi/\rho$, and the stream function is 
 \begin{equation}\label{chipm}
 \chi_{+-} = \frac{1}{2}\ln\left(\rho_+^2-2\rho_+\rho_-\cos\phi_{+-} + \rho_-^2\right)=\ln \rho_{+-},
\end{equation}
where we introduced $\phi_{+-} = \phi_+ - \phi_-$ and the convenient abbreviation
$\rho_{+-} = \sqrt{\rho_+^2-2\rho_+\rho_-\cos\phi_{+-} + \rho_-^2}$, which equals the distance $|z_+ - z_-|$ on the complex plane.
Note that $\lambda_+$ varies throughout the complex plane, {and its spatial dependence affects the dynamics of the vortex as it moves around the curved surface.}

These dynamical equations are readily generalized for an overall charge-neutral set of vortices
\begin{equation}\label{rjdot}
\dot{\bm r}_j = \frac{\hbar}{M}\,\hat{\bm n}_j\times \bm \nabla_j\left({\sum_k}'q_k\,\chi_{jk} -\frac{1}{2}q_j\,\ln\lambda_j\right),
\end{equation}
where  $j$ runs over all vortices and the primed sum omits the single term $k = j$.  Unfortunately these dynamical equations involve the local orthogonal unit vectors $\hat{\bm \theta}_j$ and $\hat{\bm \phi}_j$ which makes them less useful  in practice.  Instead, we  recast these dynamical equations in terms of the scalar variables $\theta_j$ and $\phi_j$ that determine the position of the $j$th vortex core and the stream function $\chi_{jk}$.

As a preliminary step, we rewrite Eq.~(\ref{general}) as an equation for the vector $\bm r(\theta,\phi)$ that defines the surface:
\begin{equation}\label{bmr}
\bm r(\theta,\phi) = f_1(\theta)\cos\phi\,\hat{\bm x} +  f_1(\theta)\sin\phi\,\hat{\bm y} +  f_2(\theta)\,\hat{\bm z}.
\end{equation}
Correspondingly, we have the associated unit vectors on the surface
\begin{equation}\label{unitvector}
\hat{\bm \theta} = \frac{1}{h_\theta}\,\frac{\partial \bm r}{\partial \theta},\quad \hat{\bm \phi} = \frac{1}{h_\phi}\,\frac{\partial \bm r}{\partial \phi},
\end{equation}
where $h_\theta$ and $h_\phi$ are the metric factors given in Eq.~(\ref{metric}).  In addition, the unit normal vector is 
\begin{equation}\label{normal}
\hat{\bm n} = \hat{\bm \theta}\times \hat{\bm \phi},
\end{equation}
and the three vectors ($\hat{\bm \theta}, \hat{\bm \phi},\hat{\bm n}$) constitute a right-handed orthonormal triad.
Each of these unit vectors varies continuously around the surface.

For the $j$th vortex,  Eqs.~(\ref{bmr}) and (\ref{unitvector})  show that
\begin{equation}
\dot{\bm r}_j = h_{\theta_j}\, \dot\theta_j \,\hat{\bm\theta}_j  +  h_{\phi_j}\, \dot\phi_j \,\hat{\bm\phi}_j.
\end{equation}
 The gradient vector has the familiar form 
 \beq\label{nabla}
 \bm \nabla = \frac{\hat{\bm \theta}}{h_\theta}\,\frac{\partial}{\partial \theta} + \frac{\hat{\bm \phi}}{h_\phi}\,\frac{\partial}{\partial \phi}.
 \eeq
 A combination with Eq.~(\ref{rjdot}) gives the desired scalar dynamical equations
\begin{eqnarray}\label{dottheta}
\dot{\theta}_j &=& - \frac{\hbar}{M}\,\frac{1}{h_{\theta_j}h_{\phi_j}}\,\frac{\partial}{\partial\phi_j} \left({\sum_k}'\,q_k\chi_{jk} -\frac{1}{2}\,q_j\ln\lambda_j\right),\\
\label{dotphi}\dot{\phi}_j &=&  \frac{\hbar}{M}\,\frac{1}{h_{\theta_j}h_{\phi_j}}\,\frac{\partial}{\partial\theta_j} \left({\sum_k}'\,q_k\chi_{jk} -\frac{1}{2}\,q_j\ln\lambda_j\right).
\end{eqnarray}

{The last term of Eq.~(\ref{dottheta}) makes no contribution because $\lambda_j$ is independent of $\phi_j$ for any axisymmetric surface. As a result it is easy to see that 
\beq
\sum_j \,q_j\,h_{\theta_j}h_{\phi_j}\,\dot\theta_j = -\frac{\hbar}{M}{\sum_{jk}}' q_jq_k\,\frac{\partial \chi_{jk}}{\partial\phi_j} = 0
\eeq
because $\partial \chi_{jk}/\partial \phi_j$ is odd under the interchange $j\leftrightarrow k$. Hence there is a conserved quantity
\beq\label{cons}
\Sigma = \sum_j q_j\sigma(\theta_j),
\eeq
where $\sigma(\theta) = \int d\theta\,h_\theta h_\phi$ is an integral similar to Eq.~(\ref{rhoint})  for $\ln \rho(\theta)$.  Note that this conclusion $d\Sigma/dt = 0 $ holds  for any allowed set of quantized vortices on an axisymmetric compact  surface and reflects the invariance of the surface under rotations around the $z$ axis. In other words, the dynamics depends only on the differences $\phi_{jk}$, rather than on $\phi_j$ and $\phi_k$ separately.}

{Equations (\ref{integral}) and (\ref{lambda}) now allow us to evaluate the partial  derivatives explicitly, thus expressing the dynamical equations wholly in terms of the coordinates of each vortex on an axisymmetric compact surface.  For simplicity, we focus on a single vortex dipole, where the charge neutrality requires $q_k = -q_j$, with $j =\pm$ and $q_j = \pm 1$. A detailed calculation yields the final results
\begin{eqnarray}
\label{thetajdot}
\dot{\theta}_j& = & \frac{\hbar q_j}{Mh_{\theta_j}h_{\phi_j}}\,\frac{\rho_j\rho_k\sin\phi_{jk}}{\rho_{jk}^2},\\
\label{phijdot}
\dot{\phi_j} &=& -\frac{\hbar\,q_j}{2Mh_{\phi_j}^2}\,\left(\frac{h_{\phi_j}'}{h_{\theta_j}} +\frac{\rho_j^2-\rho_k^2}{\rho_{jk}^2}\right).
\end{eqnarray} 
These equations have the significant advantage of being explicit coupled first-order differential equations for the dynamics of a  vortex dipole on an axisymmetric compact surface.  In particular, we have evaluated  all relevant partial derivatives so that no variables need to be held fixed.}

We now show that Eqs.~(\ref{dottheta}) and (\ref{dotphi}) have a simple and direct connection with the total energy of the charge-neutral set of quantized superfluid vortices on our axisymmetric surface.  In detail, we start from the total energy written in terms of the stream function $\chi_{kl}$ and the scale factor $\lambda_k$,  as derived previously in Eqs.~(33) and (34) of \cite{Guenther2020}:
\beq\label{Etot}
E = \frac{\hbar^2n\pi}{M} \left(-{\sum_{kl}}' q_kq_l\,\chi_{kl}+\sum_k q_k^2\ln\lambda_k\right),
\eeq
where the primed double sum omits the terms with $k~=~l$. {This expression for the total energy is equivalent to Eq.~(2.15) in Ref.~\cite{Dritschel2015}, here expressed in  physical variables $\chi$ and $\lambda$ instead of  mathematical ones.}
A straightforward calculation shows that the partial derivatives of $E$ yield the desired Hamiltonian equations of motion
\begin{eqnarray}
2\pi\hbar\, n\, q_j\,\dot\theta_j & = &\frac{1}{h_{\theta_j}h_{\phi_j}} \frac{\partial E}{\partial \phi_j},\label{thetaham}\\
\label{phiham}
2\pi\hbar\, n\, q_j\,\dot\phi_j & = &-\frac{1}{h_{\theta_j}h_{\phi_j}} \frac{\partial E}{\partial \theta_j},
\end{eqnarray}
with the total energy $E$ as the  effective Hamiltonian (compare Eq.~(38) from \cite{Guenther2020}) and ($\theta_j,\phi_j$) as  canonical variables. 

For any  Hamiltonian with no explicit time dependence, the conservation of energy follows almost by inspection.  Note that  
\begin{equation}
\frac{dE}{dt} = \sum_j\left(\frac{\partial E}{\partial \theta_j}\,\dot\theta_j + \frac{\partial E}{\partial \phi_j}\,\dot\phi_j\right).
\end{equation}
Use of Eqs.~(\ref{thetaham}) and (\ref{phiham}) immediately shows that $dE/dt = 0$. Thus  the vortex dynamics on  axisymmetric compact surfaces conserves  the total energy $E$, including the contribution of the spatially varying  scale factor $\lambda$ in Eq.~(\ref{Etot}) above.  As noted below Eq.~(\ref{cons}), the dynamical motion also conserves the angular momentum $\Sigma$. 

\section{Description of ellipsoids}
\label{sec:ellipsoid}

Here, we specialize to an axisymmetric  ellipsoid, {where it is convenient to use $a$ as length scale and write 
\beq \label{dsellipsoid}
ds^2 = a^2\left(h_\theta^2\, d\theta^2 +  h_\phi^2\,d\phi^2\right).
\eeq
We now have {\it dimensionless} metric factors
\beq\label{metricellipsoid}
h_\theta = \sqrt{\cos^2\theta +(b^2/a^2)\sin^2\theta},\quad h_\phi = \sin\theta.
\eeq
 Substitution of Eq.~(\ref{ellipsoid}) into Eqs.~(\ref{bmr}), (\ref{unitvector}),  (\ref{normal}) and (\ref{nabla})  gives the detailed form of the relevant unit vectors and the gradient operator, with appropriate factors of $a$ because of our dimensionless metric factors.}  
{For later reference, the Gaussian curvature $K$ of an axisymmetric ellipsoid is 
 \beq \label{Gauss}
 K = \frac{b^2}{\left(a^2 \cos^2\theta + b^2 \sin^2\theta\right)^2}
 \eeq
 with the dimension of an inverse length squared.
 
 Remarkably, there is  close connection between the Gaussian curvature $K$ and the curvature contribution $\ln\lambda$ to the energy in Eq.~(\ref{Etot}).  For an isothermal metric with scale factor $\lambda$, as  in Eq.~(\ref{planemetric}), Brioschi's formula~\cite{Gray1993} asserts that
 \beq\label{Brioschi}
-\frac{1}{\lambda^2}\, \nabla^2 \ln\lambda = -\frac{1}{\lambda^2}\,                                \frac{1}{\rho}\frac{d}{d\rho}\left(\rho\,\frac{d \ln\lambda}{d\rho}\right) = K.
 \eeq
The general form on the left-hand side simplifies to the second form for the axisymmetric  compact surfaces considered here, expressed  in terms  of the polar  variables ($\rho,\phi$).  Brioschi's  formula is effectively a nonlinear Poisson equation for $\ln\lambda$ with the  Gaussian curvature $K$ as the source.
 }
 
\begin{figure}[ht]
\begin{center}
      \includegraphics[width=3.0in]{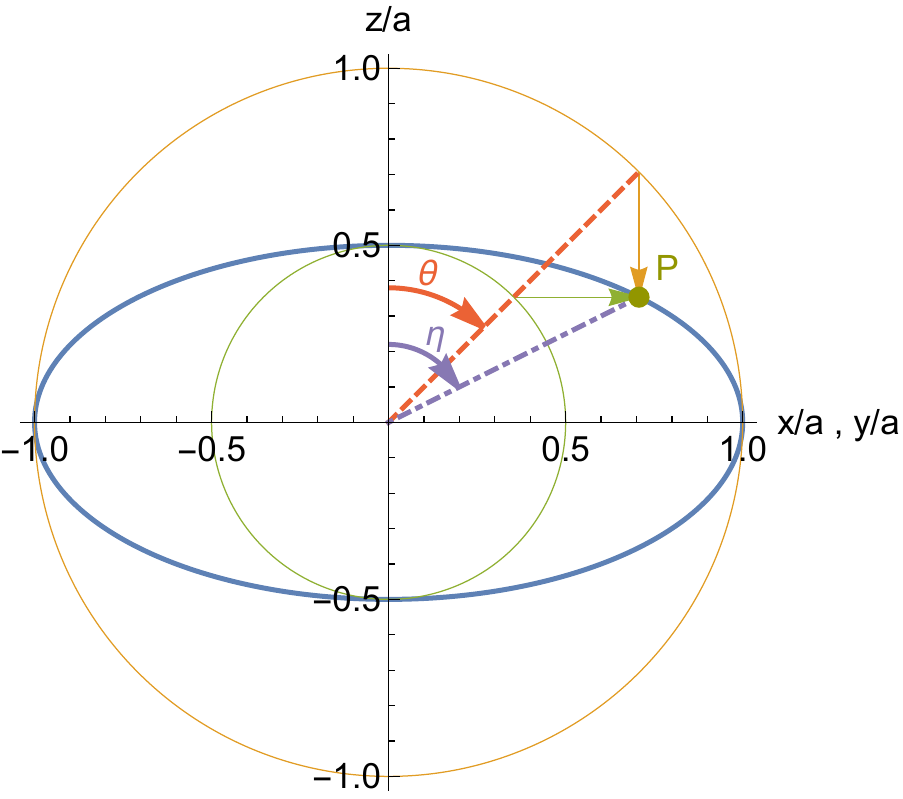}
   \caption{Cross section of an oblate ellipsoid with $b=a/2$ along with the circumscribed circle of radius $a$ and inscribed circle of radius $b$. The  vertical and horizontal intersecting lines  show de la Hire's construction with a dashed line at angle $\theta$ 
   and a  dash-dotted  line at geocentric  angle $\eta$. 
   }
   \label{fig:ellipse}
    \end{center}
 \end{figure}

Except for a sphere with $a = b$, the relation between the angular parameter $\theta$ and the polar (or ``geocentric") angle $\eta$ denoting a point on the surface of the ellipsoid is not direct, as shown in Fig.~\ref{fig:ellipse}. 
For definiteness, consider an oblate ellipsoid with $a>b$ and set $\phi = 0$, namely the right-hand side of the $xz$ plane.  Draw a circumscribed circle of radius $a$ and an inscribed circle of radius $b$.  Draw a straight line at angle $\theta$ from the $z$ axis, intersecting both circles (dashed line in Fig.~\ref{fig:ellipse}). These intersections serve to define a point P on the ellipse with $x = a\sin\theta$ and $z=b\cos\theta$.
 A straight line from the origin to this point P  (dash-dotted line in  Fig.~\ref{fig:ellipse}) defines the  ``geocentric'' polar angle $\eta$  through the equation
 \beq\label{defeta}
 \tan\eta \equiv \frac{x}{z} =\frac{a\sin\theta}{b\cos\theta} =\frac{a}{b}\,\tan\theta.
 \eeq
 
 More formally, we introduce the coordinate equations [compare Eq.~(\ref{ellipsoid})]
 \beq\label{geocentric}
 x = L(\eta)\sin\eta\cos\phi,\  L(\eta)\sin\eta\sin\phi,\  z =L(\eta)\cos\eta.
 \eeq
The latter equations  resemble those defining a sphere, with the important caveat  that the length $L$ now depends on $\eta$.
Simple algebra with Eq.~(\ref{defeta}) shows that 
\begin{eqnarray}\label{eta}
\sin\eta& =&\frac{a\sin\theta}{\sqrt{a^2\sin^2\theta+b^2\cos^2\theta}}, \nonumber\\[.2cm]
\cos\eta&= &\frac{b\cos\theta}{\sqrt{a^2\sin^2\theta+b^2\cos^2\theta}},
\end{eqnarray}
{with similar expressions based on the inverse relation $\tan\theta = (b/a)\tan\eta$.}
For a given ellipsoid and $\theta$, these relations determine the geocentric angle $\eta$. 
In addition, use of Eq.~(\ref{ellipse}) shows that
\beq\label{L}
L(\eta) = \frac{ab}{\sqrt{a^2\cos^2\eta + b^2\sin^2\eta}}.
\eeq

Before we specialize  our general formalism from Sec.~\ref{sec:formalism}  to the surface of  an axisymmetric ellipsoid, it is instructive to consider the simpler case of a sphere of radius $a$.  In this case,  Eq.~(\ref{rhoint}) simplifies to
\beq\label{rhosphere}
\rho(\theta) = \exp\left(\int\frac{d\theta}{\sin\theta}\right)=\tan(\theta/2)
\eeq
which is precisely the standard  result found with stereographic projection~\cite{Lamb1945,Bereta2021}.

For a general ellipsoid,  Eq.~(\ref{rhoint}) becomes
\beq\label{rho}
\ln\rho(\theta) = \int d\theta\,\frac{\sqrt{\cos^2\theta + (b^2/a^2)\sin^2\theta}}{\sin\theta},
\eeq
where we use Eq.~(\ref{metricellipsoid}).  The final expression depends on the details of the ellipsoid.  For the oblate (disk shaped) ellipsoid  with $a>b$, define the eccentricity $\epsilon_o = \sqrt{1-(b^2/a^2)}$ and 
\beq\label{metricoblate}
h_\theta=\sqrt{1-\epsilon_o^2\sin^2\theta}.
\eeq
The integration in Eq.~(\ref{rho}) gives %
{the result for an oblate surface}
\beq\label{rhoo}
\rho_o(\theta) =
\frac{h_\theta-\cos\theta}{\sin\theta}\,(h_\theta+\epsilon_o\cos\theta)^{\epsilon_o}.
\eeq

For the prolate (cigar shaped) ellipsoid with $b>a$, the parameter $\epsilon_o$ becomes imaginary. We define $\zeta = i\epsilon_o$ with the real quantity $\zeta =\sqrt{ (b^2/a^2)-1}$, giving
\beq\label{metricprolate}
h_\theta = \sqrt{1 +\zeta^2 \sin^2\theta}.
\eeq
The final answer 
{for a prolate surface} follows from Eq.~(\ref{rho}) 
\beq\label{rhop}
\rho_p(\theta) = \frac{h_\theta-\cos\theta}{\sin\theta}\, \exp\left[-\zeta\arcsin\left(\frac{\zeta\cos\theta}{\sqrt{1+\zeta^2}}\right)\right].
\eeq

To interpret this result, we note that a prolate ellipsoid has the 
 eccentricity $\epsilon_p = \sqrt{1-(a^2/b^2)}= \zeta/\sqrt{1+\zeta^2}$.
Equivalently, we have $\zeta =\epsilon_p/\sqrt{1-\epsilon_p^2}$.
Some algebra shows that 
Eq.~(\ref{rhop}) has the equivalent form
\begin{eqnarray}\label{rhop1}
\rho_p(\theta)&=&\frac{\sqrt{1-\epsilon_p^2\cos^2\theta}-\sqrt{1-\epsilon_p^2}\cos\theta}{
\sin\theta}\nonumber \\[.2cm]
&\times&\exp\left[-\frac{\epsilon_p}{\sqrt{1-\epsilon_p^2}}\arcsin\left(\epsilon_p\cos\theta\right)\right]
,\end{eqnarray}
where we omit an overall constant factor in the denominator that does not affect the dynamics of the vortices.

\section{Vortex dynamics on an ellipsoid}
\label{sec:dynamicsOfDipoles}

We now apply the results from Sec.~\ref{sec:dynamics} to study the dynamics of a vortex dipole 
with unit charges $q_\pm = \pm 1$ on an axisymmetric ellipsoid, with the metric given by Eqs.~(\ref{dsellipsoid}) and (\ref{metricellipsoid}).
Equation (\ref{thetajdot}) yields
\beq
\label{thetapmdot}
h_{\theta_+}\sin\theta_+\dot{\theta}_+ = h_{\theta_-}\sin\theta_-\dot{\theta}_-
= \frac{\hbar}{Ma^2}\,\frac{\rho_+\rho_-\sin\phi_{+-}}{\rho_{+-}^2
}.
\eeq
In addition, Eq.~(\ref{phijdot}) gives
\begin{eqnarray}\label{phipdot} 
\dot{\phi}_+ &=& -\frac{\hbar}{2Ma^2\sin^2\theta_+}
\left(\frac{\cos\theta_+}{h_{\theta_+}} +\frac{\rho_+^2-\rho_-^2}{\rho_{+-}^2
}\right)
\end{eqnarray}
with a similar equation for $\dot{\phi}_-$.
As noted below Eqs.~(\ref{thetajdot}) and (\ref{phijdot}), these equations are explicit first-order ordinary differential equations for the dynamics of a vortex dipole  on an axisymmetric ellipsoid.
{Finally the total energy in Eq.~(\ref{Etot}) for such a dipole reduces to 
\beq\label{Edip}
E = \frac{\hbar^2n\pi}{M}\ln\left(\rho_{+-}^2\lambda_+\lambda_-\right),
\eeq
which shows the joint role of the stream function (through the first factor) and the scale factors $\lambda_\pm$.}

\subsection{Specialize to a sphere}\label{subsec:specialize to sphere}

A sphere has $a=b$ so that  $h_\theta = 1$. Correspondingly, the radial distance on the        complex plane becomes [compare Eq.~(\ref{rhosphere})] \beq\label{rhosphere1}
\rho(\theta) = \tan\left(\frac{\theta}{2}\right)=\frac{\sin\theta}{1+\cos\theta}=\sqrt{\frac{1-\cos\theta}{1+\cos\theta}}.
\eeq
The scale factor in Eq.~(\ref{lambda}) simplifies to
\beq\label{lambdasphere}
\lambda(\theta) = \frac{\sin\theta}{\rho(\theta)}=1+\cos\theta,
\eeq
which depends on $\theta$ even though a sphere has constant Gaussian curvature.

For a sphere, it is not difficult to show that~\cite{Bereta2021}
\beq
\rho_{+-}^2
= \frac{4\sin^2(\gamma_{+-}/2)}{\lambda_+\lambda_-},
\eeq
where $\gamma_{+-} $ is the angle between $\hat{\bm r}_+$ and $\hat{\bm r}_-$. As a result, {the scale factors in Eq.~(\ref{Edip}) cancel exactly, and} the energy of a vortex dipole on a sphere has the very simple form [see Eq.~(\ref{Edip})]
\beq\label{Edipolesph}
E_{\rm dipole} = \frac{2\hbar^2 n\pi}{M}\ln\left[2\sin(|\gamma_{+-}|/2)\right],
\eeq
where $2a\sin(|\gamma_{+-}|/2)$ is the chordal distance between the two members of the vortex dipole~\cite{Lamb1945,Bereta2021}.  Hence the dynamical  motion of any dipole on a sphere conserves the chordal distance between them.  {We believe that this remarkable cancelation of the scale factors occurs only for a sphere, so that the terms $\ln\lambda_\pm$ generally play a role in the energy and the dynamics of a vortex dipole on an axisymmetric compact surface.}

The high symmetry of a sphere allows us  to consider (without loss of generality) a symmetric initial configuration relative to the equator, with $\theta_+=\theta$ and $\theta_-= \pi-\theta$ and $\phi_{+-}=0$.  Equation (\ref{rhosphere}) shows that $\rho_\pm = (1\mp\cos\theta)/\sin\theta$.  In this symmetric configuration, 
 some algebra {with Eqs.~(\ref{thetapmdot}) and (\ref{phipdot})} gives the familiar result~\cite{Lamb1945}
\beq\label{rdotsphere}
\dot{\bm r}_+ = \dot{\bm r}_-=  \frac{\hbar\hat{\bm\phi}}{2Ma}\tan\theta,
\eeq
so that the vortex dipole moves uniformly in the common $\hat{\bm \phi}$ direction along the 
great circle midway between them (the equator).
The vortices become stationary in the limit $\theta\to 0$, when they approach the north and south poles.  Near the equator, when $\theta\to \pi/2 - \delta\theta$, the translational velocity diverges, with $\dot{\bm r}_+=\dot{\bm r}_- =\hbar\hat{\bm \phi}/(2Ma\,\delta\theta)$, which is just the result for a vortex dipole on a plane since $2a\,\delta\theta$ is their spatial separation.

\subsection{Dipole placed  symmetrically around the equator }\label{subsec:symmetric equator}

As a first example of a vortex dipole on an axisymmetric ellipsoid, we consider the special configuration with the dipole symmetrically placed on either side of the equator, with $\theta_+ = \theta$, $\theta_- = \pi -\theta$, and $\phi_{+-} = 0$.  Here, the motion is purely azimuthal with 
\beq
\dot{\bm r}_+=\dot{\bm r}_- = -\frac{\hbar\hat{\bm\phi}}{2Ma\sin\theta}\left(\frac{\cos\theta}{h_\theta}+ \frac{\rho_++\rho_-}{\rho_+-\rho_-}\right),
\eeq
where
\beq\rho_\pm = \frac{h_\theta\mp\cos\theta}{\sin\theta}\,\delta_\pm.
\eeq
Here, 
\beq
\delta_\pm = \left(h_\theta\pm\epsilon_o\cos\theta\right)^{\epsilon_o}
\eeq
for an oblate ellipsoid with eccentricity $\epsilon_o$, and
\begin{eqnarray}
\delta_\pm &=& \exp\left[\mp\frac{\epsilon_p}{\sqrt{1-\epsilon_p^2}}\,\arcsin(\epsilon_p\cos\theta)\right]\nonumber \\[.2cm]
&=& \exp\left[\mp\zeta\arcsin\left(\frac{\zeta\cos\theta}{\sqrt{1+\zeta^2}}\right)\right]
\end{eqnarray}
for a prolate ellipsoid with eccentricity 
$\epsilon_p$.
For either distortion, it is not difficult to show that 
\beq
\dot{\bm r}_+=\dot{\bm r}_- = \frac{\hbar b^2\hat{\bm \phi}}{2Ma^3h_\theta}\frac{\sin\theta(\delta_++\delta_-)}{\cos\theta(\delta_++\delta_-)-h_\theta(\delta_+-\delta_-)}.
\eeq
Since $\delta_+-\delta_-$ vanishes for a sphere, this expression reduces to Eq.~(\ref{rdotsphere}) in the spherical limit.

{Figure~\ref{fig:symmdip} shows the orbits and azimuthal translational  speed $v_{\phi}(\theta)$ of a vortex dipole symmetrically placed around the equator for a sphere and for oblate and prolate ellipsoids with $a/b=3$ and $a/b=1/3$, respectively ($\epsilon_o=\epsilon_p = 2\sqrt2/3 \approx 0.943$).  The orbits themselves  are independent of the aspect ratio, but the lower panel in Fig.~\ref{fig:symmdip} shows how the speed of the dipole varies with polar angle $\theta$.

 The energy of a vortex dipole on an elongated surface depends linearly on the separation $d = |\bm r_+ -\bm r_-|$ once $d$ exceeds the radial dimension~\cite{Turner2010,Machta1989,Guenther2017}, instead of the usual logarithmic dependence when $d$ is small.   
For an ellipsoid, the translational velocity $\dot{\bm r}_+$ is proportional to $\hat{\bm n}_+\times \bm\nabla_+ E$, and the linear dependence of the energy on $d$ thus  explains the  extended flat region for a prolate ellipsoid when $d\gtrsim a$. In fact,  the translational speed  for a vortex dipole with $d\gtrsim R$ on an infinite cylinder with radius $R$ is $\hbar/(2MR)$~\cite{Guenther2017}. The  curve labeled prolate in the lower panel of Fig.~\ref{fig:symmdip} shows that the numerical value for our prolate ellipsoid agrees closely with the result for an infinite cylinder of radius $R = b/3$.                                                                                                                                                  
\begin{figure}[ht]
\begin{center}
      \includegraphics[width=3in]{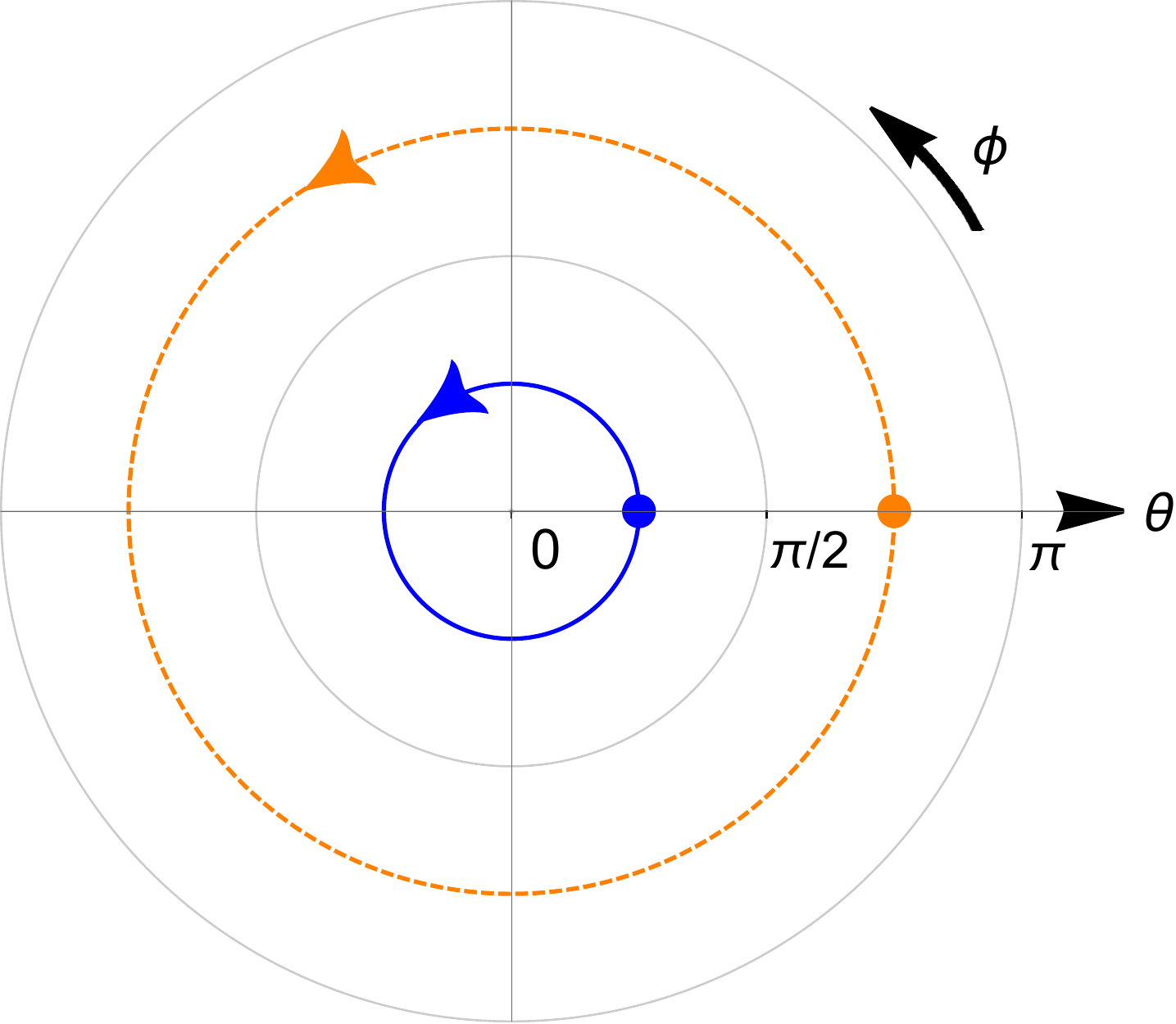}
      \vskip 0.5cm
       \includegraphics[width=\columnwidth]{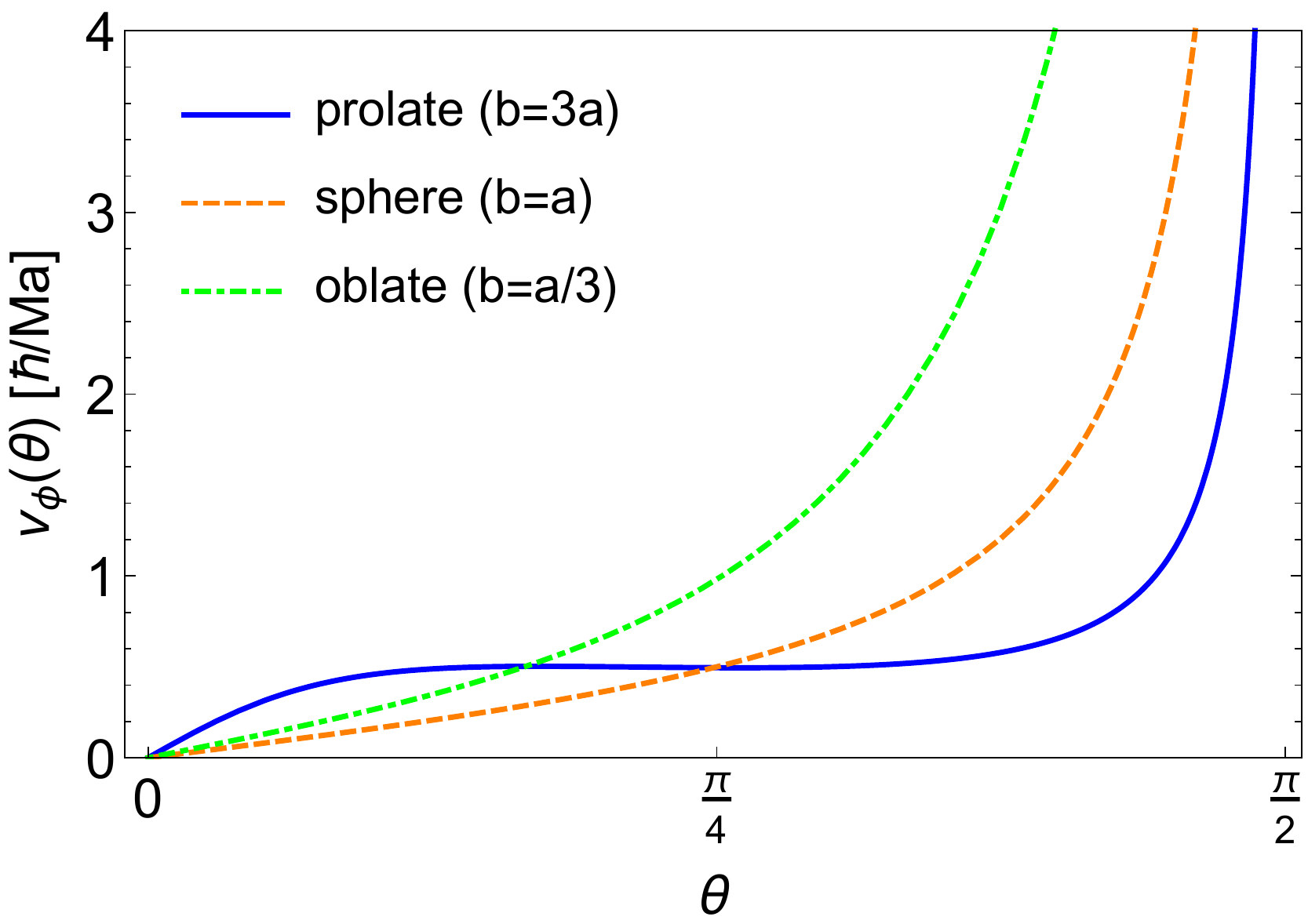}
   \caption{Upper panel:  Circulating orbits of positive (solid blue) and negative (dashed orange) member of vortex dipole placed symmetrically above and below the equator.  Lower panel: translational velocity of vortex dipole symmetrically placed around the equator for prolate and oblate ellipsoids with aspect ratios $b/a =3$ ($\epsilon_p = 2\sqrt2/3 \approx 0.943$) and $a/b=3$ ($ \epsilon_o = 2\sqrt2/3\approx 0.943$).}\label{fig:symmdip}
    \end{center}
 \end{figure}

\subsection{Energy considerations}\label{sec:energy}

The above discussion of the dynamics  of a special vortex dipole simplifies greatly  because  both vortices execute purely azimuthal motion along the common direction $\hat{\bm \phi}$.  Other initial configurations typically involve different unit vectors, and it is generally easier to start from the total energy $E_{\rm tot}$ in Eq.~(\ref{Etot}).

The simplest vortex configuration on a compact surface  is a vortex dipole with unit charges. {For this case, a combination of $\lambda_\pm = \sin\theta_\pm/\rho_\pm$ and Eq.~(\ref{Edip})} shows that the  dipole energy becomes
\begin{equation}\label{EQ}
E_{\rm dipole} = \frac{\hbar^2n\pi}{M} \ln Q,
\end{equation}
with the  conserved quantity $Q$ for  vortex dipole on an ellipsoid 
 \begin{equation}\label{Q}
Q = \left(\frac{\rho_+}{\rho_-} -2\cos\phi_{+-}  + \frac{\rho_-}{\rho_+}\right)\sin\theta_+\sin\theta_-.
\end{equation}
This expression $Q$  has several desirable features:
\begin{enumerate}
\item  $Q$ is manifestly symmetric under interchange of vortices ($\theta_+,\phi_+)\leftrightarrow (\theta_-,\phi_-)$.
\item It depends only on the combination $\phi_{+-} = \phi_+-\phi_-$ reflecting the axisymmetry of the surface.
\item  As noted in Sec.~\ref{sec:ellipsoid}, the integration for $\ln\rho$ leads to an arbitrary additive constant, and hence an arbitrary multiplicative constant for $\rho$.  Equation (\ref{Q}) shows that this constant cancels in the conserved quantity $Q$.  
\end{enumerate}

{Although Hamilton's equations are elegant,  in any particular case they involve partial derivatives with other variables held fixed.  Hence it is frequently simpler to rely on Eqs.~(\ref{thetapmdot}) and (\ref{phipdot}) that depend explicitly on the local positions, as seen  in the following Sec.~\ref{subsec:vortexatequator}.}

\subsection{Vortex dipole moving north from equator}\label{subsec:vortexatequator}

We now consider the special initial configuration of a vortex dipole with $\theta_+=\theta_- =\theta$,  starting at the equator with $\theta_0 = \pi/2$.  {This symmetry ensures that $\theta_+=\theta_-$ and therefore  $\rho_+ = \rho_- $ for all later times.}  In addition, we choose the initial value for $(\phi_--\phi_+)_0 = \Delta_0 $ to be in the range $0 < \Delta_0 < \pi$, so that the vortex dipole  starts  moving  toward the north pole. The energy  for this vortex configuration has a particularly simple form 
\begin{equation}\label{Esymm}
E = \frac{\hbar^2 n\pi}{M}\ln\left[4\sin^2\theta\sin^2\left(\frac{\Delta }{2}\right)\right],
\end{equation}
so that 
\begin{equation}\label{Qsymm}
Q= 4\sin^2\theta\sin^2(\Delta/2)
\end{equation}
 from Eq.~(\ref{Q}).  
 
  \begin{figure}[ht] 
\begin{center} 
      \includegraphics[width=3in]{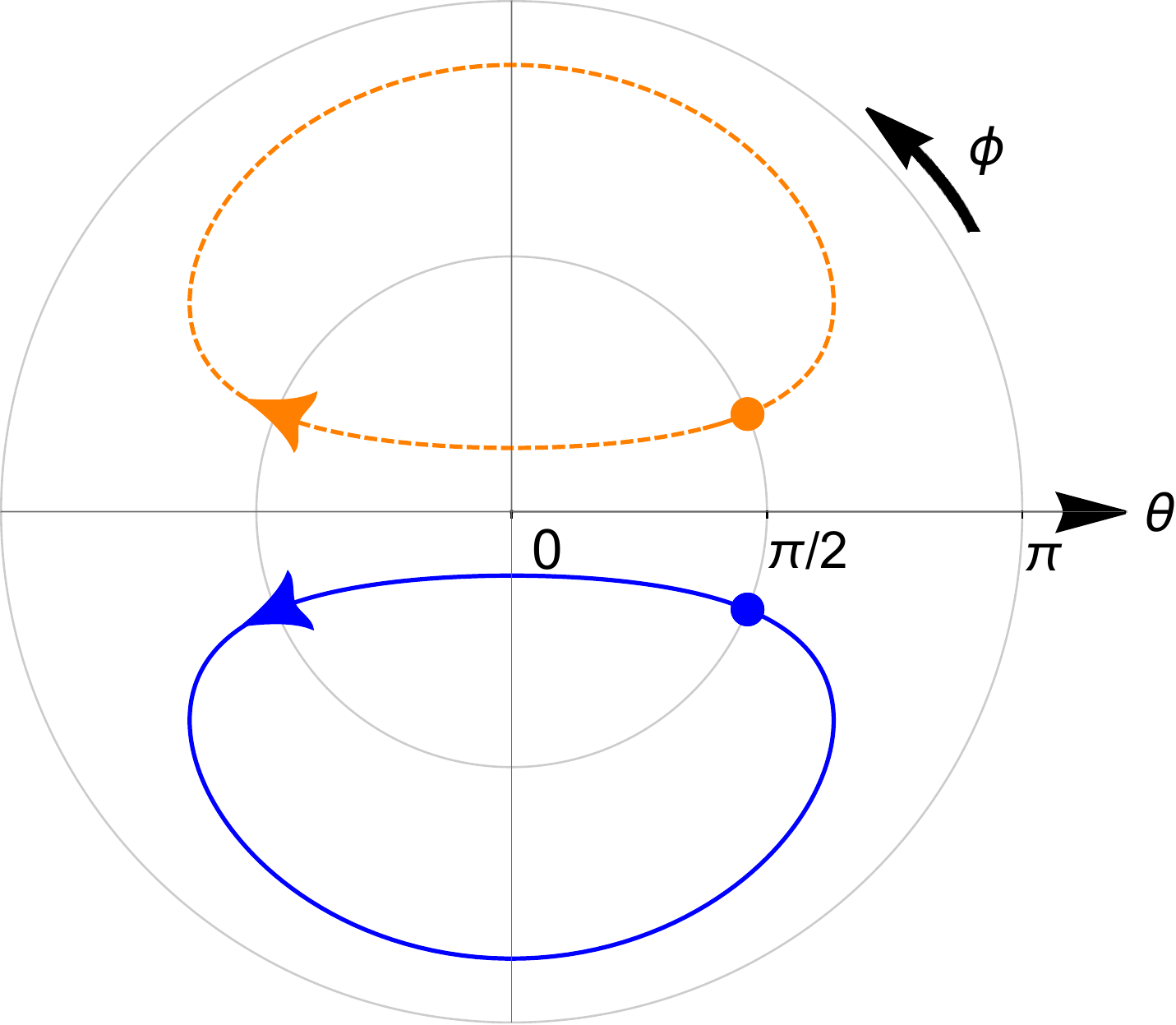}
   \caption{Circulating orbits of positive (solid blue) and negative (dashed orange)  member of vortex dipole, starting from equator $\theta = \pi/2$, marked by dots.}
   \label{fig:dipoleatequator}
    \end{center}
 \end{figure}  
 
  For these  initial conditions, $\sin\theta$ decreases from $1$ and $\sin(\Delta/2)$ increases from $\sin(\Delta_0/2)$.  Eventually $\sin(\Delta/2)$ reaches its maximum value $1$ for $\Delta  = \pi$ as the vortex dipole moves over the shoulder  at the polar angle $\theta_{\rm min} =\Delta_0/2$.  The vortex dipole then  moves back toward the equator on the opposite side of the ellipsoid,  with $\Delta$ now decreasing and  $\sin\theta$ correspondingly increasing.  Evidently, this motion will repeat cyclically, as seen in Fig.~\ref{fig:dipoleatequator}.

How does this $Q$ compare with the squared chordal distance?  
It is easy to verify that $|\bm r_+-\bm r_-|^2 = r_{+-}^2 = 4a^2\sin^2\theta\sin^2(\Delta/2) =a^2 Q$.  For this  symmetric example, we see that the dynamical motion indeed conserves the chordal distance.  {As seen below, more generic initial configurations of the vortex dipole will not conserve the chordal distance.} 

It is instructive to write out the corresponding dynamical equation for the angular variables.  Equations (\ref{thetapmdot}) and (\ref{phipdot})  together give 
\begin{eqnarray}\label{dotthetapm}
\dot{\theta}_- &= &\dot{\theta}_+= -\frac{\hbar}{2Ma^2\,h_\theta\sin\theta}\cot(\Delta/2),\\
\dot\phi_- &= &-\dot{\phi}_+  =  \frac{\hbar}{2Ma^2\,h_\theta\sin\theta} \cot\theta.\label{dotphipm}
\end{eqnarray}
As expected from the conservation of $Q$, we see that the two polar angles move together while the two azimuthal angles move  to conserve their sum.  

\subsection{General asymmetric case}
If the initial locations of the two vortices do not obey special symmetries, the ensuing dynamics becomes much more complex. One such example is shown in Fig.~\ref{fig:asymmetricorbits}. The two vortices perform a relatively regular motion, but their orbits are not closed.  
\begin{figure}[ht] 
\begin{center}
 \includegraphics[width=3in]{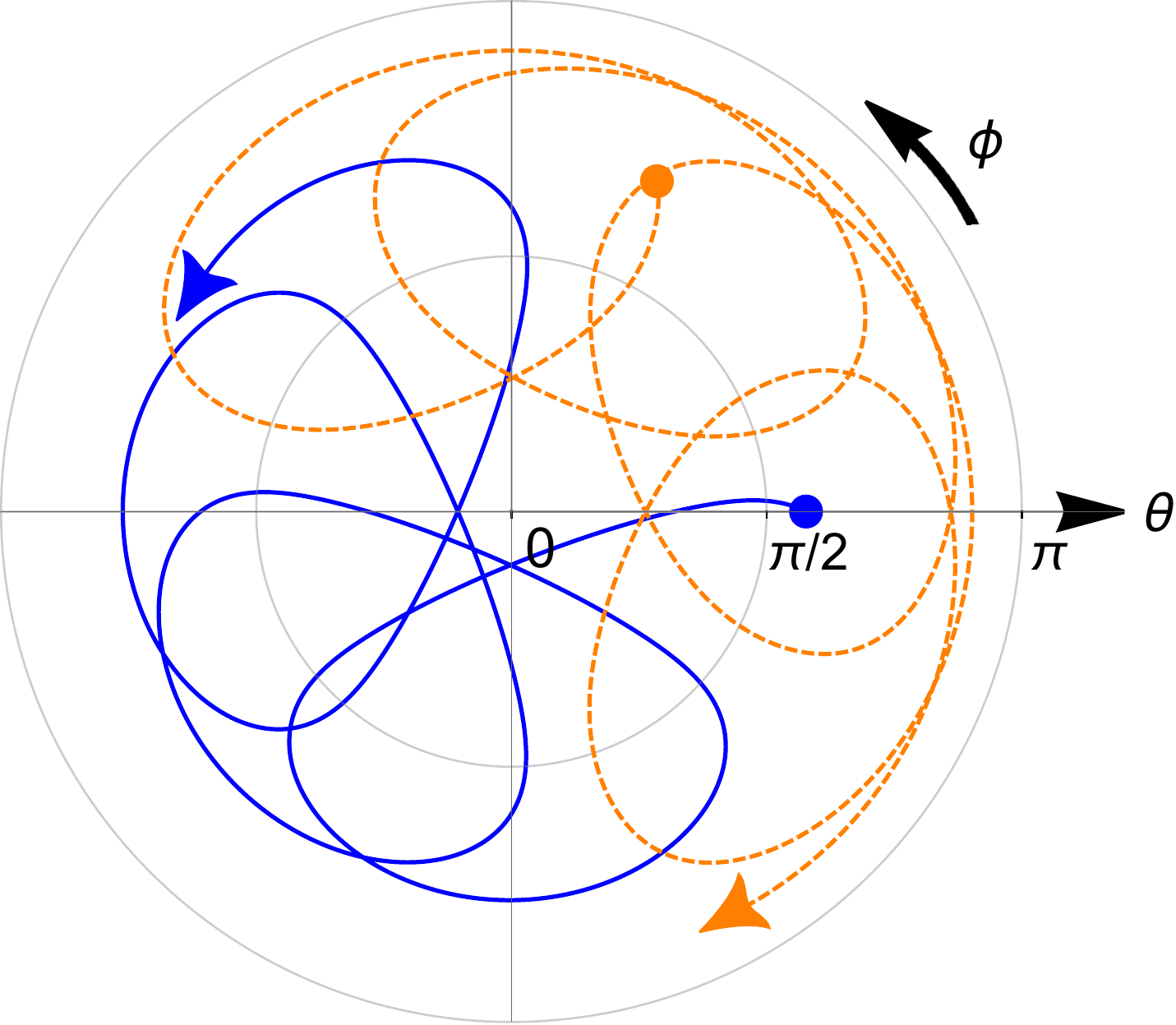}
   \caption{Circulating orbits of positive (solid blue) and negative (dashed orange) member of vortex dipole, starting from 
   two arbitrary points (marked by dots), 
   for an oblate ellipsoid with aspect ratio $b/a =1/3$ ($\epsilon_o \approx 0.943$). The vortices execute a quasi-periodic motion without forming closed orbits. See Ref.~\cite{SM} for a video displaying the simulation of this dynamics.} 
   \label{fig:asymmetricorbits}
    \end{center}
 \end{figure} 

\begin{figure}[ht!] 
\begin{center}
 \includegraphics[width=\columnwidth]{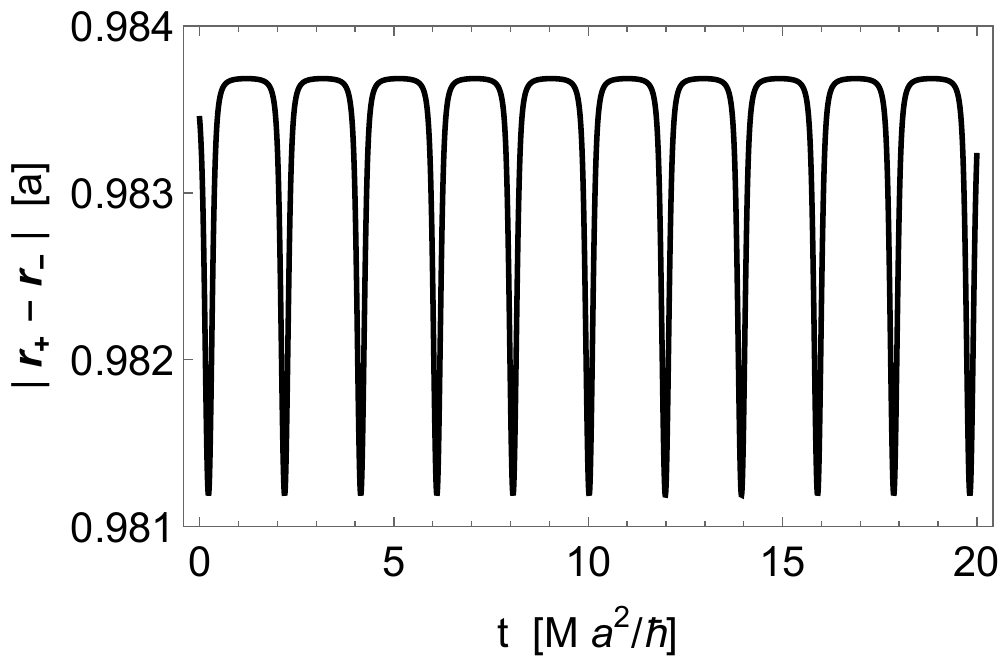} \vskip 0.5cm
  \includegraphics[width=\columnwidth]{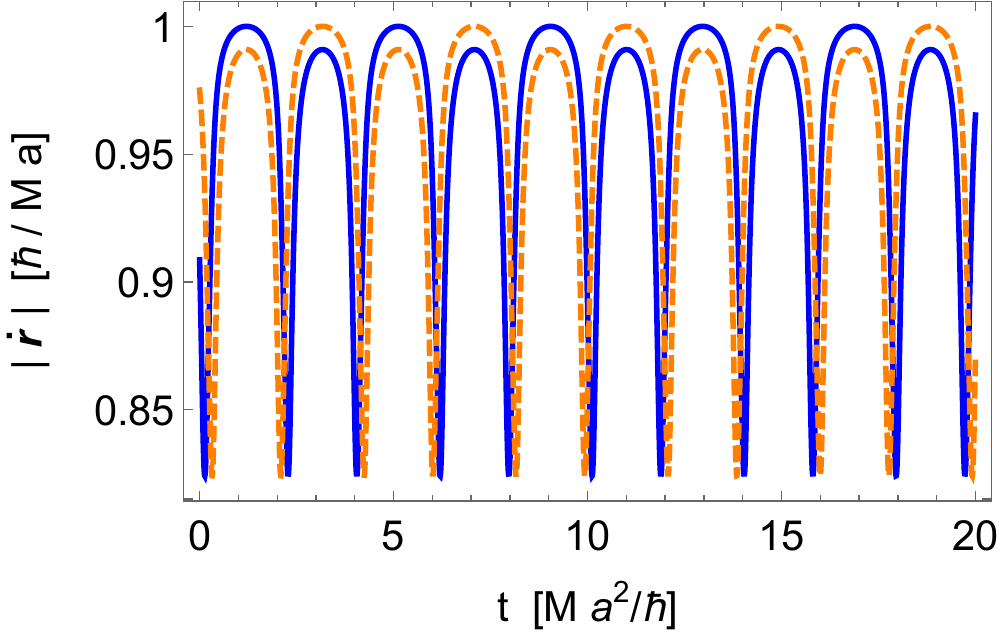}
  \vskip 0.5cm
  \includegraphics[width=\columnwidth]{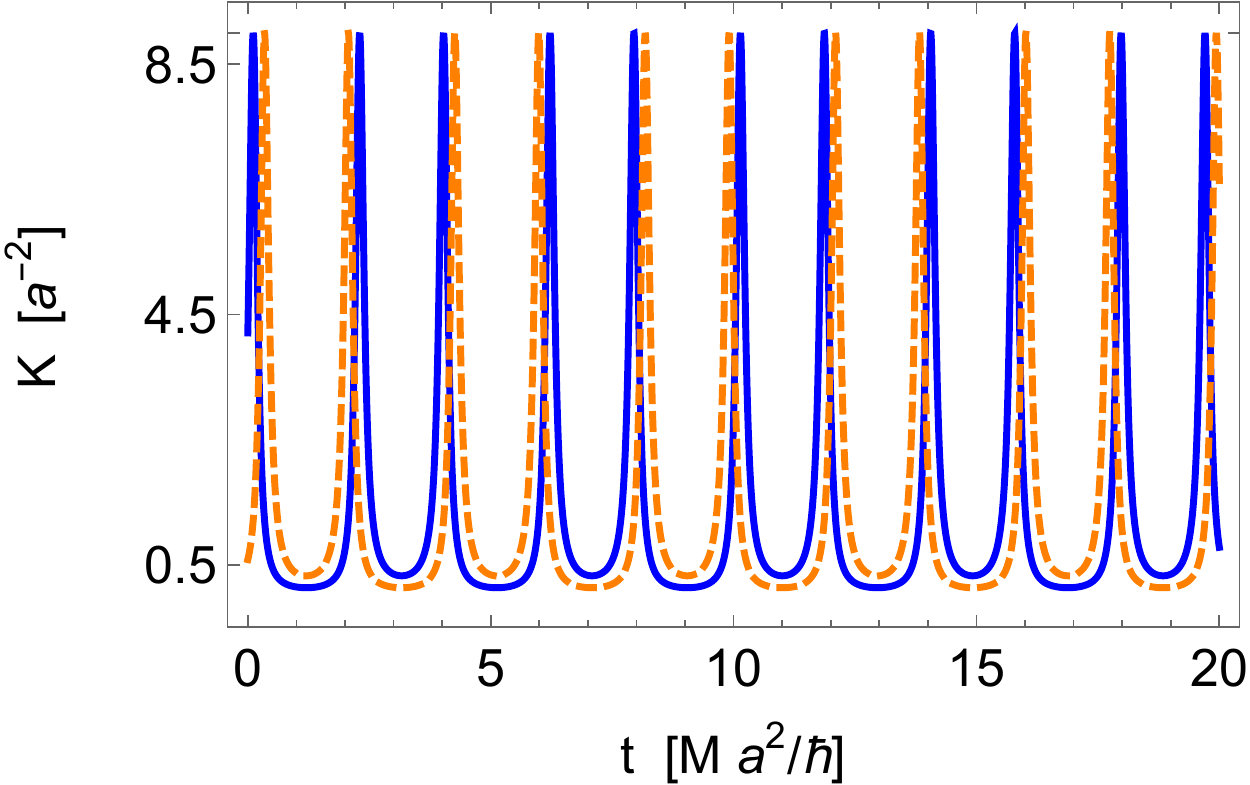}
  \caption{Upper panel: the time evolution of chordal distance; middle panel: the speed of positive (solid blue) and negative (dashed orange) vortices; lower panel: the  local Gaussian curvature for the positive (solid blue) and negative (dashed orange) vortices for the particular asymmetric configuration represented in Fig.~\ref{fig:asymmetricorbits}.}
  \label{fig:chordaldistance}
    \end{center}
 \end{figure} 
 
 We checked in detail that the dynamical motion conserves  both the total energy $E$ and the ``angular momentum'' $\Sigma$ defined in Eq.~(\ref{cons}). The upper image of  Fig.~\ref{fig:chordaldistance} shows that the chordal distance is definitely not constant, unlike our previous examples.  Instead, it  oscillates anharmonically but nonetheless periodically. The middle image shows that the speed of each vortex also varies periodically but at half the frequency of the chordal distance.  
In contrast, the sum of the squared speeds  of the two vortices $|\dot {\bm r}_1|^2 + |\dot {\bm r}_2|^2$ oscillates  
exactly in phase with (and with the same period of) the chordal distance. 
Here the squared speed is given by $|\dot{\bm r}|^2 = a^2\,h_\theta^2 \dot\theta^2 + a^2\,h_\phi^2 \dot\phi^2$. The lower panel shows the local Gaussian curvature at the position of each vortex
\begin{equation}
K  =\frac{1-\epsilon_o^2}{a^2(1-\epsilon_o^2\sin^2\theta)^2},   \end{equation}
here written for an oblate ellipsoid.
Comparison of the middle and lower panels of Fig.~\ref{fig:chordaldistance} highlights the anti-correlation between the vortex speed and the local curvature of the surface: vortices slow down in the regions of higher curvature.

\section{Conclusions}
\label{sec:conclusions}

We have constructed a theory  of quantized superfluid two-dimensional vortices on    axisymmetric compact surfaces with no holes. Specifically, we  developed a general method to transform conformally from  the axisymmetric surface  to a  plane, where we use familiar methods based on the hydrodynamic stream function.  This  transformation yields an additional term in the  dynamics of each vortex reflecting the local curvature at its position. Our approach shows the  close connection to the energy  of these vortices.  They constitute a Hamiltonian dynamical system, with the angular positions ($\theta_j,\phi_j$) as canonical variables. 

A vortex dipole is the simplest superfluid state that satisfies the condition of vanishing   total vortex charge.  We study the dynamics of two simple symmetric vortex-dipole  configurations  and then consider  a more general asymmetric initial   configuration, showing trajectories and other related quantities.  Unlike the situation on a sphere, the chordal distance is not in general a constant of the motion.

In recent years, various mathematicians have studied two-dimensional vortices on  axisymmetric  surfaces~\cite{Dritschel2015,Hally1980,Castilho2008} and triaxial ellipsoids~\cite{Regis2018}.  In particular, 
Ref.~\cite{Dritschel2015} independently developed the same method for vortices on axisymmetric surfaces based on conformal mapping.  All their examples treat classical vortices, frequently focusing on large-scale atmospheric vortices and polygonal rings of identical vortices.  In these examples, one or more polar vortices (or sometimes uniform vorticity) enforce the condition of net vortex-charge neutrality.  Another example of mathematical interest is the classical three-vortex problem on a sphere~\cite{Newton2001}, where at least two of them necessarily have different absolute vortex charges.  

The situation is quite different for the  quantized superfluid vortices that we study here.   Typically all the vortices have unit charge ($q_j=\pm 1$) and the condition of irrotational flow precludes distributed vorticity. The simplest superfluid configuration is a vortex dipole, and we studied several examples of vortex dynamics on an ellipsoid.

Reference~\cite{Guenther2020} studied superfluid vortex dynamics on a torus and other toroids of revolution,  
extending Kirchhoff's conformal transformation from a torus  to a plane~\cite{Kirchhoff1875}. It  would be interesting to study more general multiply connected surfaces (such as a pretzel) with two or more holes, where each nontrivial quantized circulation loop will contribute to the vortex  dynamics.    In such cases, however, the absence of rotational invariance presents a  definite challenge.
 
All studies of vortex dynamics on superfluid films assume a uniform thickness and density.  In practice, recent observations with the NASA Cold Atom Laboratory~\cite{Carollo2021}  (see also the theoretical prediction in Ref.~\cite{Tononi2020}) suggest that asymmetries in the apparatus will produce films with nonuniform number density.  The method of a time-dependent variational  Lagrangian has been successful for dynamical studies of nonuniform three-dimensional Bose-Einstein condensates~\cite{PerezGarcia1996}, and a similar approach should be useful for nonuniform films.

A recent study has shown that  a two-component BEC can support vortices with filled cores.  The resulting composite structure acts like a massive vortex and obeys second-order differential equations, very different from the usual first-order equation of vortex dynamics.  For example, the familiar precession of a one-component vortex in a trap can also include small-amplitude rapid oscillations around the uniform motion~\cite{Richaud2021}. It would be interesting to extend these studies to a vortex dipole on a sphere and other axisymmetric surfaces.

\section*{Acknowledgments}

M.A.C. acknowledges support by the S\~{a}o Paulo Research Foundation (FAPESP) under Grant No.~2013/07276-1. P.M. acknowledges support by the Spanish MINECO (Grants No.~FIS2017-84114-C2-1-P and PID2020-113565GB-C21), by EU FEDER Quantumcat, and by the National Science Foundation under Grant No.~NSF PHY-1748958.

\bibliography{Vortices}

\end{document}